\documentclass[12pt]{article}
\usepackage{graphics}
\usepackage{amsthm}
\usepackage{amssymb}
\usepackage{amsmath}
\usepackage{amsfonts}
\usepackage{epic}
\usepackage{hyperref}
\usepackage{epsfig}
\usepackage{bm}
\usepackage{amscd}

\theoremstyle{plain}

\theoremstyle{definition}

\def\be{\begin{equation}}
\def\ee{\end{equation}}
\smallskip

\begin{document}

\headsep=-0.5cm

\begin{titlepage}
\begin{flushright}
\end{flushright}
\begin{center}
\noindent{{\LARGE{2D quantum gravity partition function \\ on the fluctuating sphere}}}

\smallskip
\smallskip
\smallskip

\smallskip
\smallskip

\smallskip

\smallskip

\smallskip
\smallskip
\noindent{\large{Gaston Giribet and Mat\'{\i}as Leoni}}
\end{center}

\smallskip

\smallskip
\smallskip

\centerline{Physics Department, University of Buenos Aires}
\centerline{{\it Ciudad Universitaria, pabell\'on 1, Buenos Aires, 1428, Argentina.}}

\smallskip
\smallskip

\smallskip
\smallskip

\smallskip

\smallskip

\begin{abstract}
Motivated by recent works on the connection between 2D quantum gravity and timelike Liouville theory, we revisit the latter and clarify some aspects of the computation of its partition function: We present a detailed computation of the Liouville partition function on the fluctuating sphere at finite values of the central charge. The results for both the spacelike theory and the timelike theory are given, and their properties analyzed. We discuss the derivation of the partition function from the DOZZ formula, its derivation using the Coulomb gas approach, a semiclassical computation of it using the fixed area saddle point, and, finally, we arrive to an exact expression for the timelike partition function whose expansion can be compared with the 3-loop perturbative calculations reported in the literature. We also discuss the connection to the 2D black hole and other related topics. 
\end{abstract}

\end{titlepage}



\newpage


\section{Introduction}

Liouville field theory (LFT) is ubiquitous in theoretical higher energy physics. It naturally arises in the formulation of the 2D quantum gravity \cite{Seiberg, Distler} and in the path integral quantization of string theory \cite{Polyakov}. It also has applications to 3D gravity \cite{Henneaux}, to $\mathcal{N}=2$ superconformal gauge theories in 4D \cite{AGT}, to string theory on AdS$_3 \times N$ backgrounds \cite{Ribault}, to AdS/CFT holography in the stringy regime \cite{Gaberdiel}, to dS/CFT correspondence \cite{Laura}, and to many others subjects, including black hole physics and cosmology. The timelike version of theory also has interesting applications, for example in the study of dynamical tachyon condensation \cite{ST}. The consistency of timelike LFT as a {\it bona fide} CFT together with its applications to physics have been investigated by many authors in the last twenty years \cite{Gutperle:2003xf, ST, HMW, S, Zreloaded, 05, 052, KP, KP2, Giribet, McE, Schomerus:2012se, Maltz:2012zs, Picco:2013nga, Ribault:2015sxa, 190512689, 200201722, Kapec}, and in a series of recent works the theory has been revisited from a new perspective: In \cite{151105342, 151107450}, LFT was considered in the context of quantum cosmology and a Weyl-invariant formulation of gravity. In \cite{210601665}, the Euclidean path integral of 2D gravity with positive cosmological constant was studied in the semiclassical limit and its connection to timelike LFT was discussed. The problem was analyzed by expanding the theory around the two-sphere saddle point. The main idea was that, while the Euclidean 2D gravitational path integral is in general highly fluctuating, when it is coupled to a matter CFT with large central charge it admits a semiclassical two-sphere saddle point that renders the problem tractable perturbatively. After choosing a Weyl gauge, the computation of the partition function reduces to that for timelike LFT. The authors of \cite{210601665} presented results up to 2-loops and provided a path integral computation of the LFT central charge in the semiclassical limit. This led them to propose an expression for the 2D gravitational partition function on the two-sphere, which boils down to the expression of the partition function of timelike Liouville on the sphere topology. More results of this calculation appeared in \cite{210604532}, where the author reviews 2D quantum gravity coupled to a matter CFT in the fixed area ensemble and in the semiclassical limit. The computation of the partition function on the two-sphere is done in \cite{210604532} at 2-loops using the path integral approach, what is achieved by considering all the relevant Feynman diagrams and incorporating the fixed area constraint. As in \cite{210601665}, it was found that all UV divergences cancel to 2-loop order. These results were extended in \cite{220204549}, where a computation at 3-loops was performed in a similar framework. More works along this line appeared recently: In \cite{211105344}, the semiclassical gravitational path integral was discussed in relation to random matrices, and in \cite{211104715} the timelike LFT computation was performed in presence of boundaries. The latter work nicely shows how subtle the analytic continuation that connects the spacelike LFT to its timelike version can be. The subtleties of the analytic extension of LFT from spacelike to timelike signature is well known and it is clearly expressed by the computation of the 3-point correlation function on the sphere, where the timelike analog of the Dorn-Otto-Zamolodchikov-Zamolodchikov (DOZZ) formula does not agree with the {\it naive} analytic extension of the spacelike formula. That is the reason why any computation of timelike quantities that strongly relies on the spacelike formulae has to be revised and, when possible, computed separately. Here, we present a detailed computation of the Liouville partition function on the fluctuating sphere at finite values of the central charge, both for the spacelike and the timelike theory. After introducing LFT in section 2, in section 3 we discuss the derivation of the spacelike partition function from the DOZZ formula. In section 4, we rederive this quantity by using a Coulomb gas approach that turns out to be suitable to later compute the timelike partition function. The spacelike fixed area partition function is discussed in section 5, and a semiclassical computation of it using the fixed area saddle point is given in section 6. In section 7, we present the exact expression for the timelike partition function, we discuss its properties, its difference with respect to the spacelike quantity, and its relation with other results presented in the literature.

\section{Liouville field theory}

LFT is a non-compact conformal field theory \cite{Yu,Teschner} defined by the action
\begin{equation}
S_{L}[\Lambda ]=\frac{1}{4\pi }\int d^{2}z\left( \partial \varphi \bar{%
\partial }\varphi +\frac{1}{2\sqrt{2}}QR\varphi +4\pi \Lambda e^{\sqrt{2}%
b\varphi }\right) \, .  \label{mancha}
\end{equation}%
$\Lambda $ is a positive parameter that corresponds to the cosmological constant. The background charge $Q$ takes the value $Q=b+b^{-1}$ for the Liouville potential barrier $\Lambda e^{\sqrt{2}b\varphi }$ to be a marginal operator. The second term in (\ref{mancha}) involves the scalar curvature, $R$, and in the conformal gauge it has to be understood as keeping track of a $\delta$-function contribution to the curvature coming from the point at
infinity. The CFT is defined on the Riemann sphere, $\mathbb{CP}^1$. The theory is globally defined on $\mathbb{CP}^1$ once one specifies the boundary
conditions, which is done by imposing the behavior $\varphi \simeq -2\sqrt{2}Q\log |z|$ for large $|z|$. Under holomorphic transformations $z\rightarrow w(z)$ the Liouville field
transforms by acquiring an inhomogeneous piece that depends on $Q$; namely $\varphi (z) \rightarrow \varphi (w)
-\sqrt{2}Q\log |\frac{dw}{dz}|$. The central charge of the theory also depends on $Q$ and is given by $c_L=1+6Q^2$, which is bigger than $25$ provided $b$ is real.

One is typically interested in coupling LFT to another CFT, which would represent the matter content of the full theory. We will refer to this putative matter CFT as $\mathcal M$ and we will denote its central charge $c_{\mathcal{M}}$. In this way, we have the total central charge
\begin{equation*}
c=1+6Q^{2}+c_{\mathcal{M}}.
\end{equation*}%
Weyl anomaly cancellation thus demands
\begin{equation}
    Q=\sqrt{\frac{25-c_{\mathcal{M}}}{6}}\, ,
\end{equation}
which requires $b$ to be non-real for large positive $c_{\mathcal{M}}$. In the semiclassical limit ($b\to 0$) of the spacelike theory, in contrast, $Q$ is positive and large, so that $c_{\mathcal{M}}$ has to be a large negative number. 

The vertex operators of the full theory are given by \cite{Teschnervertex} 
\begin{equation*}
\mathcal{O}_{\alpha ,h}(z) = V_{\alpha }(z)\times V^{\mathcal{M}}_{h}(z)=e^{\sqrt{2}%
\alpha \varphi (z)} \times V^{\mathcal{M}}_{h}(z).
\end{equation*}%
These operators create Virasoro primaries on $\text{Liouville}\times \mathcal{M}$ of conformal dimension $\Delta =\alpha
(Q-\alpha )+h$, with $h$ being the contribution of the matter CFT; that is, $V^{\mathcal{M}}_{h}(z)$ are dimension $h$ conformal primaries of the theory ${\mathcal{M}}$. Normalizability in the spacelike theory demands $\alpha\in Q/2+i\mathbb{R}$. Here, we will focus on the LFT factor of the theory, so we will omit the matter content just assuming it is there.

\section{Partition function from DOZZ}

The LFT $N$-point correlation functions on the sphere are defined as follows \cite{Teschner,ZZ,Zreloaded,DO}
\begin{equation}
\left\langle\,  V_{\alpha
_{1}}(z_{1}) \,  V_{\alpha
_{2}}(z_{2}) \, ... \, V_{\alpha _{N}}(z_{N})\, \right\rangle_{L} \, =\, \int_{\varphi_{(\mathbb{CP}^1)}}
\mathcal{D}\varphi \,\,\, e^{-S_{L}[\Lambda ]}\, \prod_{i=1}^{N}e^{\sqrt{2}\alpha _{i}\varphi
(z_{i})}\label{normaN}
\end{equation}
where the subscript $\varphi_{(\mathbb{CP}^1)}$ indicates that the field takes values on the Riemann sphere, having imposed the boundary conditions given above: on $\mathbb{CP}^1$ the field configurations are conditioned to obey the
asymptotic $\varphi (z)\simeq -2Q\log |z|$ near $z =\infty $.

In particular, the 3-point correlation functions are given by
\begin{equation}
\left\langle V_{\alpha
_{1}}(z_1) V_{\alpha
_{2}}(z_2)V_{\alpha _{3}}(z_3 )\right\rangle _L  = \prod_{a<c}^3|z_a-z_c|^{2\Delta-4(\Delta_{a}+\Delta_{c})}\, C(\alpha_1, \alpha_2, \alpha_3)
\end{equation}
where $\Delta_a =\alpha_a(Q-\alpha_a)$, $\Delta = \sum_{a=1}^3 \Delta_a$, and where the structure constants are given by the DOZZ formula \cite{DO, ZZ}
\begin{equation}
C(\alpha_1, \alpha_2, \alpha_3) = \Big(\pi \Lambda \gamma(b^2)b^{2-2b^2}\Big)^{(Q-\alpha)/b}
\frac{\Upsilon_b(b)}{\Upsilon_b(\alpha -Q)}
\prod_{a=1}^3\frac{\Upsilon_b(2\alpha_a)}{\Upsilon_b(\alpha -2\alpha_a)} \label{CN}
\end{equation}
where $\alpha = \sum_{a=1}^3 \alpha_a$, $\gamma (x)=\Gamma(x)/\Gamma(1-x)$. The function $\Upsilon_b (x)$ was introduced in \cite{ZZ} and can be written in terms of Barnes' double $\Gamma $%
-functions $\Gamma _{2}(x|y)$ \cite{Barnes, Barnes2} as follows%
\begin{equation}
\Upsilon _{b}(x)= \Gamma _{2}^{-1}(x|b,b^{-1})\Gamma
_{2}^{-1}(b+b^{-1}-x|b,b^{-1}).
\end{equation}
This function obeys the reflection properties
\begin{equation}\label{refla}
\Upsilon_b(x)=\Upsilon_b(Q-x)\ , \ \ \ \Upsilon_b(x)=\Upsilon_{b^{-1}}(x) \ , 
\end{equation}
together with the shift properties
\begin{equation}\label{xiva}
\Upsilon_b(x+b)=\gamma (bx)b^{ 1- 2b x}\Upsilon_{b}(x) \ , \ \ \  \Upsilon_b(x+b^{- 1})=\gamma (b^{- 1}x)b^{- 1+ 2b^{- 1} x}\Upsilon_{b}(x)\, .
\end{equation}

The spacelike LFT partition function on the fluctuating sphere can be easily obtained from the 3-point function above. To do so, firstly one resorts to $PSL(2,\mathbb{C})$ invariance to fix $z_1=0$, $z_2=1$, and $z_3=\infty$ in the 3-point function; secondly, one considers the particular case $\alpha_{1}=\alpha_{2}=\alpha_{3}=b$, and, finally, one writes the relation
\begin{equation}
\frac{d^3}{d\Lambda^3}Z[\Lambda ]= \, -\, C(b,b,b)  \label{ZZZ}.
\end{equation}%
Integrating this expression and using the functional properties (\ref{refla})-(\ref{xiva}), one easily finds
\begin{equation}
Z[\Lambda ]= \frac{(1-b^{2})\left( \pi \Lambda \gamma (b^{2})\right)^{Q/b}}{\pi
^{3} Q\gamma (b^{2}) \gamma (b^{-2})},  \label{ZZZ}
\end{equation}
which is the expression of the spherical partition function of spacelike ($c_L\geq 25$) LFT. Now, let us analyze this result:
\begin{itemize}
    \item Expression (\ref{ZZZ}) is the exact result for the spacelike Liouville partition function on the
sphere topology, valid for finite $b$. It turns out to be a non trivial function of $b$ which oscillates with growing frequency and decreasing amplitude as $b^{-2}$ approaches the values $b^{2}=0$ and $b^{2}=1$. 
\item The expression for $Z[\Lambda ]$ turns out to be finite. Although it is usually said -- for example in the context of string theory-- that the partition function on the sphere topology vanishes, this is not necessarily the case for theories with no Poincar\'e invariance in the target space. More precisely, translation symmetry in the non-compact field space produces an infinite volume factor that comes to compete with the infinite volume of the conformal Killing group in the denominator of the partition function, yielding an undetermined factor of the form $\infty / \infty $. Therefore, while the partition function per unit of length might vanish, the actual partition function can in principle be finite, cf. \cite{KKK}. We will come back to this point later. 
    \item The Knizhnik-Polyakov-Zamolodchikov (KPZ) scaling of the partition function goes like $Z[\Lambda ] \sim \Lambda ^{Q/b}$, which in the semiclassical limit $b\to 0$ behaves like $\sim \Lambda^{1/b^2}$.
    \item Expression (\ref{ZZZ}) is valid for $b\in \mathbb{R}$, which means $c_L\geq 25$. The analogous formula for $c_L\leq 1$ corresponds to the timelike partition function and it will be given below, in section 7. 
    \item One can set the cosmological constant $\Lambda $ to an arbitrary positive value $\Lambda _*$ by shifting $\varphi \to \varphi - \frac{1}{\sqrt{2}b}\log ({\Lambda}/{\Lambda_*})$. In particular, one can choose the special value $\Lambda _*=\gamma(1-b^2)/\pi $, which makes the entire factor $(\pi \Lambda \gamma(b^2))^{Q/b}$ to disappear from the expression.
    \item In the limit $b^2 \to 1$ ({\it i.e.} $c_L\to 25$) the function $Z[\Lambda_*]$ diverges. This is related to the fact that 2D string theory with $c_{\mathcal{M}}=1$ requires a renormalization of $\Lambda$, cf. \cite{Verlinde}.
    \item At $b^2\in \mathbb{Z}_{\geq 2}$ the function $Z[\Lambda_*]$ also diverges, exhibiting a double pole; similarly for $b^{-2}\in \mathbb{Z}_{\geq 2}$.
    \item One of the intriguing features of the expression (\ref{ZZZ}) is the fact that it does not exhibit the
Liouville self-duality under the transformation $b\rightarrow 1/b$. Typically, correlation functions of LFT are invariant under $b \to 1/b$ provided one also transforms the cosmological constant as $\Lambda \to (\pi \Lambda \gamma(b^2))^{1/b^2}/(\gamma(b^{-2})\pi )$; this is why the breakdown of the self-duality at the level of the partition function is puzzling. This has been observed by Al. Zamolodchikov in \cite{perturbed} and it will be explained below in the Coulomb gas formalism. 
\item Before concluding this section, a word on normalization: For that of $Z[\Lambda]$ to be a sensible computation, its normalization -- or, at least, the $b$-dependent part of it-- has to be determined somehow naturally. Here, we have considered the canonical normalization that follows from (\ref{normaN}) and that is consistent with the DOZZ formula (\ref{CN}).
\end{itemize}

\section{Coulomb gas computation of the partition function}

As we have just seen, the derivation of the spherical partition function from the DOZZ formula is quite direct and succinct: literally, one page! However, such a derivation is not possible in the timelike theory, at least not without assuming unjustified aspects of the analytic continuation. Therefore, being interested as we are in presenting a method by means of which the analytic continuation from spacelike to timelike is under control, here we will give an alternative derivation of (\ref{ZZZ}). Such a method is given by the Coulomb gas approach, cf. \cite{gastonmati}. This amounts to consider the free field form of the correlator; namely
\begin{align}
\left\langle \, V_{\alpha _{1}}(z_{1})\,  ...\, V_{\alpha _{N}}(z_{N})\, \right\rangle _L  = \frac{\Lambda ^{s}\Gamma (-s)}{b}  \int_{\mathbb{C}^s} \prod_{t=1}^{s} d^{2}w_{t} \, \int_{\varphi_{(\mathbb{CP}^1)}} \mathcal{D} \varphi \, e^{-S_L[0]}\, \prod_{i=1}^{N}e^{{\sqrt{2}\alpha
_{i}}(z_{i})}\, \prod_{r=1}^{s}  e^{{\sqrt{2}b}(w_{r})}
\label{despues}
\end{align}%
This follows from expanding  the interaction term $\Lambda e^{\sqrt{2}b\varphi }$ and integrating the zero mode of $\varphi $. The integration over the zero mode produces a factor $\int dx\, e^{-\Lambda x}x^{-1-s}=\Lambda ^{s}\Gamma
(-s)$ together with a $\delta $-function $b^{-1}\, \delta ( s+b^{-1}(\alpha _{1}+\alpha _{2}+...\alpha
_{N})-1-b^{-2})$ fixing $ s=-b^{-1}\sum_{i=1}^N\alpha _{i}+1+b^{-2}$, and selecting in this way a single term in the expansion, {\it i.e.} the term of order $\Lambda ^s$, which comes with the insertion of $s$ integrated interaction operators $V_{b}(w_r)=e^{\sqrt{2}b\varphi (w_r)}$, $r=1, 2, ... \, s$. The latter operators, being marginal in the free ($\Lambda=0$) CFT, can be thought of as the screening operators required by the presence of a background charge $Q$. These operators also admit the string theory interpretation of being the vertices of the tachyons of which the Liouville barrier (or wall) is made. In this picture, the cosmological constant $\Lambda $ is related to the string coupling constant, the Liouville wall is what prevents the theory to reach the strong coupling regime, and the background charge $Q$ governs the slope of the lineal dilaton: The background charge enters in the integration over the zero mode of $\varphi $, yielding
\begin{equation}
bs+\sum_{i=1}^{N}\alpha _{i}=Q,\label{pupo2}
\end{equation}%
where the Gauss-Bonnet theorem has been used on the right hand side of (\ref{pupo2}), keeping in mind that we are formulating the theory on the sphere: $\frac{1}{4\pi}\int dz^2 R = \frac 12 \chi (  \mathbb{CP}^1  )=1$.

Correlators (\ref{despues}) can be computed by performing Wick contractions of the $N+s$
operators, using the free field propagator $\left\langle \varphi (z_{1})\varphi
(z_{2})\right\rangle =-2\log |z_{1}-z_{2}|$. The latter yields the operator product expansion $e^{\sqrt{2}\alpha
_{1}\varphi (z_{1})}e^{\sqrt{2}\alpha _{2}\varphi (z_{2})}\sim
|z_{1}-z_{2}|^{-4\alpha _{1}\alpha _{2}}e^{\sqrt{2}(\alpha _{1}+\alpha _{2})\varphi
(z_{2})}+...$. Here, we are interested in the partition function; that is to say $N=0$. In that case, the number of screening operators to be integrated out turns out to be $m=s-3=b^{2}-2$ as the volume of the
conformal Killing group $PSL(2,\mathbb{C})$ should be cancelled by fixing three
out of the $s$ local operators $e^{\sqrt{2}b\varphi (z)}$. We use projective invariance to insert three of these operators at the points $z_{1}=0$, $z_{2}=1$ and $z_{3}=\infty $. (This has to be distinguished
from the direct computation of the 3-point function \cite{ZZ} of three
light states $\alpha _{1}=\alpha _{2}=\alpha _{3}=b,$ where one has to integrate over the insertion points of all the $s$ screening operators and not over $s-3$ of them).

The latter discussion is related to something we have said before: while the partition function per unit of length might vanish, the actual partition function can in principle be finite \cite{KKK}. In theories such as LFT, where the exponential potential ({\it i.e.} the wall) breaks target space translation invariance, the exact volume factor can be computed by the trick we have just introduced: fixing three screening operators and integrating the zero mode of $\varphi $ explicitly.

Another important remark about (\ref{despues}) has also to do with the number of integrals to be performed: $s-3$. In the case of the partition function this is given by $s=1+b^{-2}$, which in general is not a positive integer but a real number. Therefore, expression (\ref{despues}) has to be taken formally: In order to define the theory for arbitrary $b\in \mathbb{R}$ one has to analytically extend that expression. A natural way of doing so is to first assume $s\in \mathbb{Z}_{\geq 1}$ and then extend the integrated expression to the domain $s\in \mathbb{R}$. This strategy was shown to reproduce the correct results in a variety of CFTs. The free field
techniques as the one considered here were developed by Dotsenko and Fateev 
\cite{Fateev2,Dotsenko} in the context of Generalized Minimal Models and later extended by Goulian and Li \cite{GLi} to LFT; see also \cite{dot1,dot2,D}. Using these techniques, the partition function $Z[\Lambda ]$ is given by
\begin{equation}
Z[\Lambda ]=\frac{\Lambda ^{m+3}}{b}\Gamma (-m-3)
\int_{\mathbb{C}^m} \prod_{t=1}^{m} d^{2}w_{t}\int_{\varphi_{(\mathbb{CP}^1)}} \mathcal{D}\varphi \, e^{-S_{L}[0]}\, e^{\sqrt{2}b\varphi (0)}e^{\sqrt{2}b\varphi (1)}e^{\sqrt{2}b\varphi
(\infty )}\prod_{r=1}^{m}e^{\sqrt{2}b\varphi (w_{r})}  \label{empieza}
\end{equation}%
with $m=-2+b^{-2}$. The vertex inserted at infinity has to be understood as being accompanied with the appropriate factor that extract the singularity, namely in the $\lim_{z_{3}\rightarrow \infty
}|z_{3}|^{-4} e^{\sqrt{2}b\varphi
(z_{3})}$. After applying the Wick rules, we get%
\begin{equation*}
Z[\Lambda ]=b^{-1}\Lambda ^{3+m}\Gamma (-m-3)\int_{\mathbb{C}^m} \prod_{i=1}^{m} d^{2}w_{i}\left(
\prod_{r=1}^{m}|w_{r}|^{-4b^2 }|1-w_{r}|^{-4b^2
}
\prod_{t=1}^{m}
\prod_{\ell =1 }^{t-1}
|w_{t}-w_{\ell }|^{-4b^2 }\right) ,
\end{equation*}%
which can be explicitly solved for integer $m$ by using the Dotsenko-Fateev
formula for generalized Selberg type integrals; see (B.9) in \cite{Dotsenko}. As explained above, while we are
interested in the case where $m$ is generic, in order to solve the integral expression we first assume that $m$ is a positive integer and then extend the final
expression. Assuming $m\in \mathbb{Z}_{\geq 1}$, we get%
\begin{equation*}
Z[\Lambda ]=\frac{\Lambda ^{3+m}}{b}\Gamma (-m-3)\Gamma (m+1)\pi ^{m}\gamma
^{m}(1+b^2 )\prod_{r=1}^{m}\gamma (-rb^2 )\gamma
^{2}(1-(r+1)b^2 )\gamma (-1+(r+2+m)b^2 ).
\end{equation*}%
where $\gamma (x)=\Gamma (x)/\Gamma (1-x)$. The non-trivial task here is to perform the analytic continuation in $m$. In
order to do this, we can rewrite the expression above by using that $\gamma (-1+(r+2+m)b^2 )=\gamma (rb^2 )$; so, we can write
\begin{equation}
Z[\Lambda ]=\frac{\Lambda ^{3+m}}{b}\Gamma (-m-3)\Gamma (m+1)\pi ^{m}\gamma
^{m}(1+b^2 )\prod_{t=1}^{m}\gamma (-tb^2 )\gamma (tb^2
)\prod_{r=2}^{m+1}\gamma ^{2}(1-r b^2 ).  \label{introducidas}
\end{equation}
Next, further simplifications are needed: We use that $m=b^{-2}-2$, {\it i.e.} $1-rb^2 =(m+2-r)b^2 $, to arrange the last product in (\ref{introducidas}); and then we rewrite the product as follows
\begin{equation*}
\gamma (1-2b^2 )\gamma (1-3b^2 )...\gamma (1-mb^2 )\gamma (1-(m+1)b^2
)=\gamma (b^2 )\gamma (2b^2 )...\gamma ((m-1)b^2 )\gamma (mb^2 );
\end{equation*}
that is
\begin{equation*}
\prod_{r=2}^{m+1}\gamma (1-rb^2 )=\prod_{r=1}^{m}\gamma (rb^2 ).
\end{equation*}%
Then, we use $\gamma (-rb^2 )\gamma (1+rb^2 )=1$ to write%
\begin{equation}
Z[\Lambda ]=b^{-1}\Lambda ^{1+b^{-2}}\Gamma (-m-3)\Gamma (m+1)\pi ^{m}\gamma ^{m}(1+b^2
)\gamma ^{2}(b^2 )(-1)^{m}b ^{-4m}\Gamma ^{-2}(m+1), \label{trombamarina}
\end{equation}
where the identities $\gamma (x)\gamma (-x)=\gamma (x)\gamma ^{-1}(1+x)=-x^{-2}$
were used. The properties of the $\gamma $-function are also used
to write $\gamma (2-b^{-2})=-(1-b^{-2})^{2}\gamma (1-b^{-2})$, $\gamma (1+b^{2} )=-b^{4}\gamma (b^{2} )$, and $\gamma (-1+b^{2} )=-(1-b^{2}
)^{-2}\gamma (b^{2} )$, and finally, once all is written in terms of $b,$ the
partition function takes the form
\begin{equation}
Z[\Lambda ]=\frac{(1-b^{2})\left( \pi \Lambda \gamma (b^{2})\right) ^{1+b^{-2}}}{\pi
^{3}(b+b^{-1})\gamma (b^{2})\gamma (b^{-2})},  \label{ZZZ2}
\end{equation}%
which exactly reproduces (\ref{ZZZ}). This is the exact result for the spacelike Liouville partition function, which here we have obtained by the Coulomb gas formalism using free field techniques. 

In order to fully appreciate this calculation of $Z[\Lambda ]$, it is
worth comparing it with
the analogous computation of the structure constant $C(\alpha _{1},\alpha _{2},\alpha _{3})$ for the particular
configuration $\alpha _{1}=\alpha _{2}=\alpha _{3}=b$. In particular, this comparison permits to understand the breakdown of Liouville self-duality. The main difference
between the calculation of $Z[\Lambda ]$ and that of $C(b,b,b)$ is given by the overall factor $\Gamma (-s)=\Gamma
(-m-3)$ in (\ref{empieza}). As said, this factor comes from the
integration over the zero-mode of $\varphi $, but it can be also
thought of as coming from the combinatorial problem of permuting all the
screening operators: Actually, for integer $s$ this factor can be written as 
$\Gamma (-s)\sim (-1)^{s}\Gamma (0)/s!$, where the infinite factor $\Gamma (0)$
keeps track of a divergence due to the non-compactness of the Liouville
{\it direction}. This yields the factorial $1/s!$ arising in the residue
corresponding to the poles of resonant correlators. In
the computation of the structure constant $C(b,b,b)$, in contrast to that of $Z[\Lambda ]$, such overall factor is $\Gamma (3-s)$
instead of $\Gamma (-s)$ since one has to divide by the permutation of $s-3$
screening charges instead. Therefore, we find 
\begin{equation}
    C(b,b,b)\, =\, \frac{\Gamma (3-s)}{\Gamma
(-s)} \, \frac{Z[\Lambda ]}{\Lambda ^3}=-\frac{s!}{(s-3)!} \, \frac{Z[\Lambda ]}{\Lambda ^3}\, =\, (1+b^{-2})b^{-2}(1-b^{-2}) \, \frac{Z[\Lambda ]}{\Lambda ^3}, 
\end{equation}
which is consistent
with the relation $\frac{d^{3}Z}{d\Lambda ^{3}}=-C(b,b,b)\sim \Lambda ^{b^{-2}-2}$ we used in section 3, cf. \cite{ZZ}. In other words, this combinatorial problem appears to be at the root of the breakdown of the Liouville self-duality at the level of the
partition function.

\section{Fixed area partition function}

Now, let us study the partition function at fixed area. The motivation to do so is, on the one hand, to explore the semiclassical limit of this quantity; on the other hand, it will enable us to crosscheck the normalization of our result (\ref{ZZZ}): We already made the point that, for the computation of the partition function to be meaningful, it is better that the normalization gets naturally fixed.   

Instead of
considering the usual partition function $Z[\Lambda ]$, we now consider the fixed area
partition function $Z^{(A)}$, with the area being defined
as $A=\int d^2z e^{\sqrt{2}b\varphi}$. The two partition functions are related each other as follows
\begin{equation}
{Z}[\Lambda]=\int_{\mathbb{R}_{\geq 0}}{Z}^{(A)}\ e^{-\Lambda A}\
\frac{dA}{A}.\label{trafe}
\end{equation}
From the KPZ scaling of $Z[\Lambda ]$, it can be shown that\footnote{Compare the sign of this formula with the one appearing in \cite{ZZ} and with the one appearing in \cite{perturbed}.}
\begin{equation}
{Z}^{(A)}=\frac{1}{\Gamma(-Q/b)}(A\Lambda)^{-Q/b}{Z}[\Lambda].
\end{equation}
Using (\ref{ZZZ}) and using properties of the $\Gamma$-function, we have
$Q\gamma(b^{-2})\Gamma(-Q/b)=b^3\Gamma(b^{-2})$, and then we get the following expression for the fixed area partition function
\begin{equation}\label{quantum}
{Z}^{(A)}=\left(\frac{A}{\pi}\right)^{-Q/b}
\frac{(1-b^2)[\gamma(b^2)]^{1/b^2}}{(\pi b)^3 \Gamma(b^{-2})}.
\end{equation}
Some comments are in order:
\begin{itemize}
    \item The fixed area partition functions scales as $Z^{(A)}\sim A^{-1-b^{-2}}$, which by definition reproduces the correct KPZ scaling $Z[\Lambda ]\sim \Lambda^{1+b^{-2}}$ after transforming as in (\ref{trafe}); that is, 
    \begin{equation}
        \frac{1}{\Gamma(-1-b^{-2})}\int _0^{\infty}  e^{-\Lambda A}A^{-2-b^{-2}}\, dA\, =\, \Lambda^{Q/b}.
    \end{equation}
    \item Studying the analytic properties of the $\hat{Z}[\Lambda]$ amounts to take into account the zeroes it exhibits at $b^2\in \mathbb{Z}_{\geq 1}$ due to the factor $(\Gamma(1-b^2))^{-1/b^2}$. 
    \item The behavior for small $b$ is given by the following expansion
    \begin{equation}
        Z^{(A)}\simeq \left(\frac{A}{\pi}\right)^{-1-b^{-2}}
\, \frac{e^{-2b^{-2}\log b+\mathcal{O}(1)}}{b^3\pi ^3} \, \Big( 1 - b^2  +\mathcal{O}(b^5)\Big)
    \end{equation}
\item The semiclassical limit, which corresponds to small $b$, yields the following leading exponential behaviour
\begin{equation}\label{semi}
{Z}^{(A)}\simeq e^{1/b^2}\left(\frac{A}{\pi}\right)^{-1/b^2}.
\end{equation}
Below, we will show that this semiclassical behavior is exactly reproduced by the saddle point and, besides, that it is consistent with the AdS$_3$/CFT$_2$ correspondence.  
\end{itemize}

\section{Semiclassical partition function}

\begin{figure}
\centering
\includegraphics[scale=0.50]{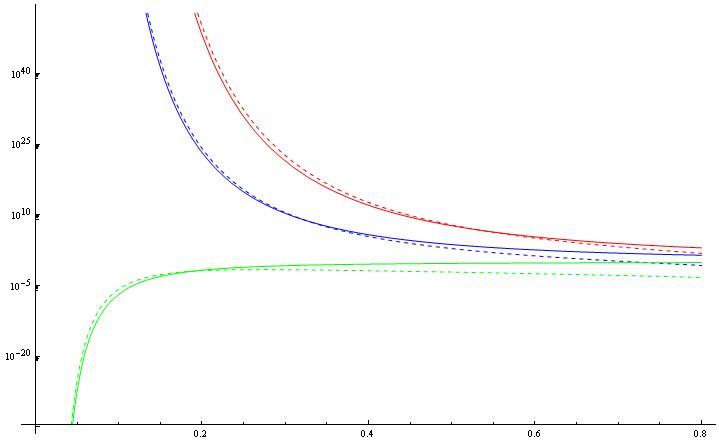}
\caption{Comparison between the classical and the quantum partition function: The dashed lines are the exact result (\ref{quantum}) while the solid lines are the saddle point approximation (\ref{semi2}). The horizontal axis is $b$, while different values of $A$ correspond to curves of different colors: $A=0.1$ (red), $A=1$ (blue), $A=10$ (green). This manifestly shows that the semiclassical approximation matches the exact result for small $b$ and for all values of $A$.} \label{fig:1}
\end{figure}

Now, let us study the semiclassical theory. The classical limit of LFT corresponds to take $b \to 0$. This is achieved by conveniently rescaling $\varphi \to \varphi /(\sqrt{2}b)$ and $\Lambda \to \Lambda /(4\pi b^2)$ in (\ref{mancha}); cf. \cite{Higher}. We are interested in making an estimation of the
fixed area partition function using the saddle point approximation
\begin{equation}
{Z}^{(A)}\simeq \exp{\left(-\frac{1}{b^2}S_A^{(cl)}\right)},
\end{equation}
where $S_A^{(cl)}$ is the constant area classical Liouville action defined by
\begin{equation}
S_{L}[\varphi]=S_A^{(cl)}[\varphi]+\Lambda b^2\int d^2z\,e^{\varphi} \, ;
\end{equation}
see equations (2.34)-(2.39) in \cite{ZZ}. To find the classical solution needed for the evaluation of the classical constant
area action, we have to solve the positive curvature Liouville equation
\begin{equation}
\partial\bar\partial\varphi=\frac{2\pi}{A}\left(\sum\eta_i-1\right)e^{\varphi},
\end{equation}
with the usual boundary conditions, where $\eta_i$ are the classical momenta. Since we are involved with the
partition function, we take $\eta_i=0$. Up to a
global conformal transformation, there is
a unique, everywhere regular solution with the right asymptotics; namely
\begin{equation}
\varphi_0=\log\left(\frac{A}{\pi(1+z\bar z)^2}\right).
\end{equation}
We insert this solution in the regularized constant area classical action, and we
perform the integral around a big circle of radius $R$, taking into account the bulk
action, the boundary term, and the counterterm. We get
\begin{equation}
S_A^{(cl)}[\varphi_0;R]=\log
R^2+\frac{1}{R^2}-1+\log\left(\frac{A}{\pi}\right)-2\log(1+R^2)+\log R^2 \, ,
\end{equation}
so that in the limit of $R\rightarrow\infty$ we have
\begin{equation}
S_A^{(cl)}[\varphi_0]=-1+\log\left(\frac{A}{\pi}\right).\label{HHHabove}
\end{equation}
Therefore, the saddle point approximation tells us that
\begin{equation}\label{semi2}
{Z}^{(A)}\simeq e^{-\frac{1}{b^{2}} S_A^{(cl)}[\varphi_0]}=
e^{1/b^2}\left(\frac{A}{\pi}\right)^{-1/b^2},
\end{equation}
in perfect agreement with the semiclassical leading behaviour of the quantum
partition function found in (\ref{semi}); see Figure \ref{fig:1} for a detailed comparison.

An interesting property of the semiclassical result (\ref{semi2}) is that it also matches the expectation from AdS$_3$/CFT$_2$ correspondence. In reference \cite{Henneaux}, which has been regarded as one of the precursors of AdS/CFT correspondence, it has been shown that LFT appears as a semiclassical description of the boundary dynamics of 3D gravity in asymptotically AdS space, where the central charge of the boundary CFT$_2$ is given by $c_L= 1+6Q^2 \simeq 6b^{-2} = \frac{3\ell }{2G}$, with $\ell $ and $G$ being the AdS radius and the Newton constant respectively; see \cite{Donnay} for details. In \cite{Krasnov}, Krasnov studied this correspondence between Liouville CFT and Einstein gravity in AdS$_3$ by evaluating the Einstein-Hilbert action $\frac{1}{16\pi G}\int d^3x\sqrt{-g}(R+2/\ell^2)$ on configurations that correspond to puncture surfaces and comparing it with corresponding LFT correlators. In equation\footnote{Eq. (39) in the preprint.} (3.12) of \cite{Krasnov} the semiclassical expression for the 3D gravity action with no punctures is given, and it exactly matches the result (\ref{HHHabove}) above.

\section{Timelike partition function}

Now, let us consider the timelike theory. Timelike LFT is defined starting from the spacelike theory and performing the Wick rotation $\varphi \to -i\varphi $ along with $b \to ib$. These changes suffice to produce a real action for a field $\varphi $ with the {\it wrong} sign kinetic; namely 
\begin{equation}
 \hat{S}_{L}[\Lambda ]=\frac{1}{4\pi }\int d^{2}z\left( - \partial \varphi \bar{%
\partial }\varphi +\frac{1}{2\sqrt{2}}\hat{Q}R\varphi +4\pi \Lambda e^{\sqrt{2}%
b\varphi }\right)  \label{manchaa}
\end{equation}%
where $b$ and $\Lambda $ are two real parameters, and $\hat{Q}=b-b^{-1}$. It is customary to add a hat to the quantities and parameters of the timelike theory, {\it e.g.} the timelike structure constant will be denoted by $\hat{C}(\alpha_1 \alpha_2, \alpha_3 )$. (It is also usual to denote $\beta = ib$; we do not use this notation here, but it is convenient to take it into account when comparing with the formulae in the literature). 

Whether (\ref{manchaa}) defines an actual conformal field theory is a question that has been addressed by many authors \cite{HMW, S, Schomerus:2012se, Ribault:2015sxa}. In \cite{HMW}, the authors showed that a particular proposal for the timelike
LFT 3-point function, originally presented in \cite{Zreloaded, 05, 052, KP, KP2}, can actually be computed
by the original LFT path integral evaluated on a new integration cycle. In \cite{Giribet}, a
Coulomb gas derivation of the timelike 3-point function was shown to reproduce the same expression. This showed that the Coulomb gas approach suffices to compute timelike quantities provided the analytic extension is done appropriately. 

The timelike 3-point function takes the form 
\begin{equation}\label{TC}
\hat{C}(\alpha _{1},\alpha _{2},\alpha _{3})=\left( \pi \Lambda \gamma
(-b^{2})b^{2+2b^{2}}\right) ^{(\hat{Q}-\alpha )/b}\frac{\Upsilon
_{b}(\hat{Q}+b-\alpha )}{b\Upsilon _{b}(b)}%
\prod_{i=1}^{3}\frac{\Upsilon _{b}(2\alpha
_{i}-\alpha +b)}{\Upsilon _{b}(b-2\alpha _{i})}
\end{equation}
with $\hat{Q}=b-b^{-1}$, $\alpha = \alpha_1+\alpha_2+\alpha_3$. The most salient feature of the timelike 3-point function (\ref{TC}) is that it does not agree with the {\it naive} analytic extension $b\to ib$ of the DOZZ formula; on the contrary, it is more like its {\it inversion}. This introduces a problem when trying to compute the timelike LFT partition function, as evaluating $\alpha_1=\alpha_2=\alpha_3=b$ in (\ref{TC}) yields a vanishing result due to three $\Upsilon_b(0) = 0$ factors in the numerator: Timelike DOZZ formula $\hat{C}(b,b,b)$ identically vanishes. On the other hand, computing the timelike $\hat{Z}[\Lambda ]$ from the naive analytic continuation of the spacelike DOZZ does not seem to be well justified. All this represents a problem to compute the timelike partition function, and therefore we need to resort to a different kind of computation. This is were Coulomb gas computation becomes important \cite{Giribet}. We can compute the timelike partition function as we did in section 4 with the spacelike partition function. Following the same steps we find that in the timelike case the number of integrated screening operators would be $\hat{s}-3=b^{2}-2$. Then,
assuming $\hat{s}\in \mathbb{Z}_{>3}$, the timelike partition function reads
\begin{equation}
\hat{Z}[\Lambda ]=\frac{\Lambda ^{\hat{s}}}{b}\Gamma (-\hat{s})\Gamma
(\hat{s}-2)(-\pi \gamma (- b^{2})b^{4})^{\hat{s}-3}\prod_{r=1}^{\hat{s}-3}\frac{\gamma ( rb^{2})\gamma (-
(\hat{s}+r-1)b^{2}-1)}{\gamma ^{2}(- (1+r)b^{2})}.\label{trombamarina2}
\end{equation}
Noticing that $1+ rb^{2}= b^{2}(\hat{s}-1-r)$ we can
rearrange the products of $\Gamma $-functions and eventually find
\begin{equation}
\hat{Z}[\Lambda ]=\frac{(1+b^{2})\left( \pi \Lambda \gamma (-b^{2})\right) ^{\hat{Q}/b}}{\pi ^{3}\hat{Q}\gamma (-b^{2})\gamma (-b^{-2})},  \label{0pf}
\end{equation}
recall $\hat{Q}=b-b^{-1}$. This is the exact resut for the timelike LFT on the sphere topology, cf. \cite{Giribet}. Some comments are in order:
\begin{itemize}
   \item The result we obtained for the timelike partition function $\hat{Z}[\Lambda] $ is related to the spacelike $Z[\Lambda] $ in such a way that, if we perform $b\to ib$ then $Z[\Lambda] $ transforms into $-i\hat{Z}[\Lambda ]$, as naively expected from the path integral definition. In other words, the expression (\ref{0pf}) is obtained from the spacelike partition function by replacing $
Q\rightarrow \hat{Q}$ and $b^{2}\rightarrow -b^{2}$, up to an $i$ factor. 
\item  The expression for $\hat{Z}[\Lambda ]$ we obtained is non-zero despite $\hat{C}(b,b,b)=0$.
    \item  As in the spacelike case, the expression $\hat{Z}[\Lambda ]$ is not
invariant under Liouville self-duality $b\rightarrow 1/b$.
    \item Expanding our formula for small $b$, we get 
    \begin{eqnarray}
    \hat{Z}[\Lambda ] \simeq (\pi \Lambda )^{1-b^{-2}}\, \frac{\sin(\pi b^{-2})}{\pi^3}\, e^{-b^{-2}(2+i\pi +2\log b)-2\gamma -\frac 23 \zeta (3)b^4 +\mathcal{O}(b^6)} \, \Big(  2b^{-1} + \frac{13}{3} b + \frac{169}{36}b^3 + \mathcal{O} (b^4)\Big) \nonumber
    \end{eqnarray}
    where $\gamma =0.5772156...$ and $\zeta(3)=1.2020569...$ are the Euler-Mascheroni and the Ap\'ery constants, respectively.
    \item There is a factor $\Gamma (-b^{-2}) (\Gamma(-b^2))^{1/b^2} $ in the denominator of (\ref{0pf}), so that when $b^2\in \mathbb{Z}_{ \geq 1}$ the expression has zeroes. 
    \item It is possible to compare our result (\ref{0pf}) with a conjecture made in \cite{210601665} about the form of the timelike partition function; see equation (5.1) therein, as well as equations (3.24), (4.8)-(4.9) and section 4.2. The formula for the timelike partition function at finite $b$ proposed in \cite{210601665} was inferred from the direct extension of the spacelike DOZZ formula. A result similar to (\ref{0pf}) is obtained in this way, except for some extra factor $e^{\hat{Q}^2(1-2\log 2)+i\frac{\pi}{2}}$.
    \item A perturbative 2-loop computation of the sphere timelike partition function was also addressed in \cite{210601665}, and it was compared with the small $b$ expansion of the aforementioned conjectured formula for $\hat{Z}[\Lambda ]$. Up to normalization, the authors found a mismatch in a factor $(1-e^{{2\pi i}/{b^{2}}})$ between the perturbative and the non-perturbative result, which they interpret as being related to an extra saddle point that, unlike the loop calculation, the DOZZ formula does {compute}. Understanding the relation with the loop computation requires further study. 
    \item As usual in non-rational CFT computations, the Coulomb gas approach used in this section and in section 4 requires analytic extensions of products and sums. We have already explained how to deal with such formal expressions and with their analytic extensions; for instance, we explained the case in which the upper limit of a product is in general a non-integer real number. In some models this can actually be even more subtle as upper limits of the products are often negative integers; {\it e.g.} see (\ref{a}) below. Let us elaborate on this: The function $P(n)=\prod_{r=1}^{n}f(r)$, which is obviously well defined for $n\in \mathbb{Z}_{>0}$, can be extended for $n\in\mathbb{Z}_{\leq 0}$ by simply defining $P(-n)=\prod_{r=0}^{n-1}f^{-1}(-r)$. By exponentiation, this implies that a sum of the form $S(n)=\sum_{r=1}^n f(r)$ admits the extension $S(-n)=-\sum_{r=0}^{n-1} f(-r)$. This prescription has already been used in the context of non-rational CFTs \cite{D, Rado} and its consistency can be checked by considering simple examples such as the geometric series $\sum_{r=1}^n p^r$, the Faulhaber’s numbers $\sum_{r=1}^n r^p$ and generalizations. This leads to expressions like $\prod_{r=-n}^{+n} \gamma (rb^2)= b^{-4n-2}\gamma(-n)$ which often appear in CFT calculations. The techniques employed throughout this paper are consistent with these formulae.
    \item Before concluding, a few words on black holes: The analytic extension of the Coulomb gas realization discussed in this section and in section 4 has led to the exact formula for the LFT partition
function. Then, it is natural to ask whether a
similar computation can be done for other examples of non-compact CFTs. For instance, one could address the case of string theory on the 2D black hole \cite{Witten, Maldal, Pakmandice, DVV}, whose worldsheet CFT corresponds to the $SL(2,\mathbb{R})_{k}/U(1)$
Wess-Zumino-Witten (WZW) model, the WZW level $k$ being related to the black hole curvature in string units. This model is remarkably similar to LFT, especially if one resorts to the Wakimoto free field
representation of the $\hat{sl}(2)_{k}$ symmetry algebra \cite{Wakimoto}. The latter involves a scalar field $\varphi $ with background charge $b$ and a commuting ghost $\beta $-$\gamma $ system. In this case, the perturbative parameter -- {\it i.e.} the analog of $\Lambda $ in LFT-- turns out to be the black hole mass, $M$. The KPZ scaling of partition function is expected to be linear in $M$. Coulomb-gas like prescription to compute
correlation functions in the 2D black hole was worked out in 
\cite{B}. As in LFT, to compute the partition
function it is necessary to insert three screening operators at fixed
points. Then, the partition function turns out to be given by the following integral
integral
\begin{eqnarray}
\mathcal{I} \, =\, M^{m+3}\Gamma (-m-3)\int_{\mathbb{C}^m}\prod_{i=1}^{m}
d^{2}u_{i}\prod_{r=1}^{m}|u_{r}|^{-4b^{2}}|1-u_{r}|^{-4b^{2}}\, 
\prod_{t=1}^{m} \prod_{l=1}^{t-1} |u_{t}-u_{l}|^{-4b^{2}}
\end{eqnarray}
with $b^{-2}=k-2$. As before, this expression is merely formal since the
integration over the zero mode of $\varphi $ demands $m=-2$. However, as
before, we can treat this as holding for a positive integer number $m$ and
then try to extend the final expression for generic values of $m$. This
yields
\begin{equation}
\mathcal{I} \, =\, M\Gamma (-m-3)\Gamma (m+1)\pi ^{m}\gamma ^{m+1}(1+b^{2})\gamma
(b^{2})(-1)^{-m}\prod_{r=1}^{m}(1-rb^{2})^{-2}.  \label{a}
\end{equation}%
However, unlike in LFT, here there is an obvious obstruction in working
out this expression. In this case, the number of integrated screening operators turns out
to be $m=-2$ (instead of $m=-2+b^{-2}$ as in LFT) and thus the expression diverges as 
\begin{equation}
\mathcal{I} \, = \, -\, M\, \Gamma ^{2}(-1) \, \frac{(1+b^2)^2}{\pi^2b^4} \, = \, \infty     
\end{equation}
One of the two divergent factors $\Gamma (-1)$, the one coming from the factor $\Gamma (-m-3)$ in (\ref{a}), is typically associated to the non-compactness of the target
space \cite{diFK}, while the other, coming from the factor $\Gamma (m+1)$ in (\ref{a}), turns
out to be associated to the fact that the black hole partition function on
the sphere is proportional to the black hole mass $M$, and thus no
derivative $\frac{d^{3}Z}{dM^{3}}$ can yield a finite 3-point function. To understand this point better, let us be reminded of the fact that
the KPZ scaling black hole partition function is expected to go like $Z[M]\sim M$,
which follows from standard arguments in the path integral approach. On
the other hand, the 3-point function of three screening operators goes like 
\begin{equation}
\langle \beta (0)\beta (1)\beta (\infty )\rangle \times c.c.\times \langle e^{-\sqrt{2}b\varphi (0)}e^{-\sqrt{2}b\varphi (1)}e^{-\sqrt{2}b\varphi
(\infty )}\rangle  \, \sim \, M^{-2}\, . \label{3bbb}
\end{equation}
where $c.c.$ stands for the complex conjugate counterpart. The difference between the calculation of $Z[M ]$ in the black hole theory and of $Z[\Lambda ]$ in LFT
is ultimately due to the difference of the background charges in both
theories: Since no singular terms appear in the contractions among $\beta $ fields in (\ref{3bbb}) then such a 3-point function simply
reduces to a LFT 3-point function but with the {\it wrong}
background charge $b$ instead of $b+1/b$. This leads to a crucial difference, which boils down to the factor $\prod_{r=1}^{m}(1-rb^{2})^{-2}$ in (\ref{a}), cf. (\ref{trombamarina}).
In the case of LFT (with the correct background charge) such factor is instead $\prod_{r=1}^{m}r^{-2}=(-1)^{-4m}\Gamma ^{-2}(m+1)$; see the last factor in (\ref{trombamarina2}). In  (\ref{trombamarina}) the $\Gamma $-functions in the denominator combine with the overall factor $\Gamma
(-m-3)\Gamma (m+1)$ in the numerator, yielding $\frac{\gamma (-m)}{(1-b^{-2})b^{-2}(1+b^{-2})}$. This makes LFT computation special; no analogous computation seems possible for the 2D black hole in the continuous approach. In \cite{KKK}, a matrix model computation of the 2D black hole partition function on the sphere topology was given. This is based on the sine-Liouville FZZ dual model. The Coulomb gas computation in the FZZ dual model does correctly reproduce the KPZ scaling of that computation \cite{Tonga}.
    \end{itemize}

\section{Conclusion}

Expressions (\ref{ZZZ2}), (\ref{quantum}) and (\ref{0pf}) are the important results, which correspond to the exact formulae for the sphere topology spacelike partition function, the fixed area partition function, and timelike partition function, respectively. These have been obtained by means of different methods and have shown to pass a variety of consistency checks.

\[\]

The work of G.G. has been supported by CONICET and ANPCyT through grants PIP-1109-2017, PICT-2019-00303.

\end{document}